\documentstyle[prl,aps,epsf,floats]{revtex}
\newcommand{\beq}{\begin{equation}}
\newcommand{\eeq}{\end{equation}}
\newcommand{\beqr}{\begin{eqnarray}}
\newcommand{\eeqr}{\end{eqnarray}}

\newcommand{\w}{{\omega}}

\def\bk{{\mathbf k}}

\newcommand{\sigmab}{\mbox{\boldmath $\sigma $}}

\def\half{{1\over2}}

\def\eqa{\begin{eqnarray}}
\def\eea{\end{eqnarray}}
\def\cE{{\cal E}}

\def\cT{{\cal T}}
\def\ve{{\varepsilon}}
\def\a{{\alpha}}
\def\b{{\beta}}

\def\t{{\theta}}

\begin{document}
\draft \flushbottom \twocolumn[
\hsize\textwidth\columnwidth\hsize\csname
@twocolumnfalse\endcsname
\title{
Diamagnetic Persistent Currents and
Spontaneous Time-Reversal Symmetry Breaking in Mesoscopic Structures
}
\date{\today}
\author{Damir Herman$^1$, Harsh Mathur$^1$, and Ganpathy Murthy$^2$
 }
\address{
$^1$Physics Department, Case Western Reserve University, Cleveland, OH 44106-7079\\
$^2$Department of Physics and Astronomy, University of Kentucky,
Lexington KY 40506-0055.}
\maketitle

\begin{abstract}
Recently, new strongly interacting phases have been uncovered in
mesoscopic systems with chaotic scattering at the boundaries by two of
the present authors and R. Shankar. This analysis is reliable when the
dimensionless conductance of the system is large, and is
nonperturbative in both disorder and interactions. The new phases are
the mesoscopic analogue of spontaneous distortions of the Fermi
surface induced by interactions in bulk systems and can occur in any
Fermi liquid channel with angular momentum $m$. Here we show that the
phase with $m$ even has a diamagnetic persistent current (seen
experimentally but mysterious theoretically), while that with $m$ odd
can be driven through a transition which spontaneously breaks
time-reversal symmetry by increasing the coupling to dissipative
leads.

\end{abstract}
\vskip 0.3cm ]

The  interplay of disorder and interactions is a rich
source of unexplained phenomena in the bulk, especially in two
dimensions\cite{2dmit}, despite three decades of theoretical
effort\cite{int+disorder,belitz,weak-ferro}. In mesoscopic systems one
is confronted with phenomena not seen in the bulk, such as Coulomb
Blockade oscillations\cite{CB,CB-expt} of the zero-bias conductance,
or persistent currents in mesoscopic rings in a small external
magnetic
field\cite{persist0,persist2,persist-expt0,persist-expt0.5,persist-expt1,persist-expt1.5,persist-expt-gaas,persist-expt2}(for
reviews see\cite{mesoscopics-review}). It has been realized in the
last several years that one needs to take interactions seriously in
order to understand the experimental Coulomb Blockade peak spacing
statistics\cite{CB-expt}. This understanding has led to Universal
Hamiltonian\cite{univ-ham1,univ-ham2} treatments for weak-coupling
(electron gas parameter $r_s=1/a_0\sqrt{\pi\rho}\approx 1$, where
$a_0$ is the Bohr radius and $\rho$ is the electron density), in which
a constant charging interaction and a constant exchange interaction
are kept in addition to the single-particle energy.

Recently, two of us investigated\cite{qd-us1} the stability of the
Universal Hamiltonian to other interactions of the Landau Fermi-liquid
type\cite{agd}, which are expected to be present in ballistic quantum
dots with chaotic boundary scattering, but are not in dots deep in the
diffusive limit\cite{altshuler-aronov}. The Landau interaction
$u(\t(\bk)-\t(\bk'))$ depends only on the angle between the momenta
$\bk,\bk'$ of the interacting particles\cite{agd}, and can be
parameterized in two dimensions by a set of Landau parameters $u_m$:
\beq u(\t-\t')=u_0+\sum\limits_{m=1}^{\infty} u_m \cos{m(\t-\t')} \eeq
The values of the Landau parameters depend on the strength and range
of interactions, and can only be accessed
numerically\cite{kwon-ceperley} for large $r_s$ (strong
interactions). It was found\cite{qd-us1} that while the Universal
Hamiltonian was stable in the regime $u_m>u_m^*=-1/2\ln2$, it became
unstable to a mesoscopic Pomeranchuk\cite{pomeranchuk} phase for
$u_m<u_m^*$. Soon afterwards one of us and R. Shankar\cite{qd-us2}
showed how to construct the large-$N$ theory of the strong-coupling
regime (the dimensionless conductance $g$ plays the role of $N$). Details of
this treatment will appear soon\cite{in-prep}.

In this Letter, treating Coulomb Blockade and persistent currents
within the same approach, we show that the mesoscopic Pomeranchuk
phases display unexpected signatures in the persistent current,
including a diamagnetic persistent current (seen
experimentally\cite{persist-expt1.5,persist-expt-gaas,persist-expt2}
but so far unexplained) in a model without superconductivity for $m$
even, and spontaneous time-reversal symmetry breaking for $m$ odd.

We want the effective Hamiltonian in an energy window of width the
Thouless energy $E_T$ around the Fermi energy. This is the regime of
validity\cite{mesoscopics-review} of Random Matrix Theory
(RMT)\cite{RMT}. For ballistic structures $E_T=\hbar v_F/L$, where
$v_F$ is the Fermi velocity and $L$ is the linear system size. The
dimensionless conductance is defined as the number of single-particle
energy states (of mean spacing $\Delta$) in this window
$g=E_T/\Delta$.  Our effective Hamiltonian\cite{qd-us1,qd-us2} has a
noninteracting part representing the chaotic scattering at the walls,
and a Fermi-liquid-like interacting part which conserves momentum. The
order parameter\cite{qd-us2} $\sigmab$ of the Pomeranchuk phase is a
two-dimensional vector whose magnitude $\sigma$ and direction $\chi$
represent the size and direction of the maximum Fermi surface
distortion. The shape of the deformed Fermi surface is given by
$\sigma \cos(m\theta-\chi)$, where $m$ is the angular momentum channel
in which the instability occurs.  To determine the behaviour of
$\sigmab$ the fermions are integrated out and an effective action
is obtained\cite{qd-us2}, the dominant part of which is self-averaging for large $g$. 
\beqr 
{\bar S}_{eff}=&g^2
\int dt  \bigg({|\sigmab|^2\over 2\Delta}\bigg({1\over |u_m| }-{1\over|u_m^*|}\bigg)+
\lambda (\sigmab^2)^2\bigg)+ \nonumber\\
&g\int {d\w\over2\pi} |\sigmab(\w)|^2 f(\w)
\label{eff-action-s0} 
\eeqr 

The ``kinetic'' term $f(\w)$ behaves like $\w^2$ for $\w\ll\Delta$ and
like $|\w|$ for $\w\gg\Delta$, indicating the Landau
damping of $\sigmab$. For large
$g$ the saddle-point of this action dominates\cite{qd-us2}, since
fluctuations are down by $1/g$. A strong enough attractive Landau
parameter $u_m\le u_m^*$ leads to
symmetry-breaking\cite{qd-us1,qd-us2}.

We have numerically evaluated the effective potential for the
collective variable $\sigmab$ and analyzed its dependence on external
magnetic flux. This effective potential can be obtained as the ground
state energy $\cE(\sigmab)$ of a noninteracting fermion Hamiltonian
where $\sigmab=\sigma(\cos\chi{\hat i}+\sin\chi{\hat j})$ appears as a
parameter: 
\beq (H_{\sigma}(\chi))_{\a\b}=\ve_\a \delta_{\a\b}-g\sigma
M_{\a\b}(\chi) 
\eeq 
where the first term encodes the chaotic scattering (with the
eigenvalues $\ve_\a$ controlled by RMT), and $M_{\a\b}(\chi)$
represents the coupling between the collective mode $\sigmab$ and
particle-hole excitations of the fermions\cite{qd-us2}.  In the
large-$g$ limit, the ground state energy of the system in the
strong-coupling regime is\cite{qd-us2} simply the value of the
effective potential at its global minimum. This energy automatically
contains the contributions which are specific to the particular
disorder realization in addition to the self-averaging contributions
determnined earlier\cite{qd-us2}. The averaged energy landscape of
$\sigmab$ in the strong-coupling phase would be a ``Mexican Hat'',
with rotational symmetry\cite{qd-us2}. The sample-specific
contributions break this  symmetry completely for $m$ even,
leading to a single minimum, and to a two-fold symmetry for $m$ odd,
as can be seen in Figure 1.
\begin{figure}
\narrowtext
\epsfxsize=2.4in\epsfysize=1.6in
\hskip 0.3in\epsfbox{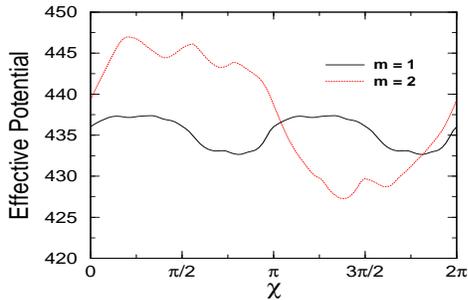}
\vskip 0.15in
\caption{The effective potential in the Mexican Hat in units of $\Delta$ 
plotted as a function of the angle $\chi$ for the cases $m$ even and
odd for $g=20$ and $u_m=1.7$. }
\label{evschi}
\end{figure}

To see the relation of the above to the persistent current we note
that\cite{mesoscopics-review} \beq I(\phi)=-{\partial
F(\phi)\over\partial\phi}.  \eeq where $\phi$ is the external flux
piercing the sample.  At zero temperature the Free energy $F$ is just
the ground state energy $\cE$, so we desire to obtain $\cE$ as a
function of $\phi$. We obtain this by taking the noninteracting part
of the Hamiltonian from the RMT ensemble of crossover
Hamiltonians\cite{RMT} parametrized as
\beq
H_{cross}=\sqrt{1\over{1+C^2\phi^2}}\bigg(H_{S}+C\phi H_{A}\bigg)
\eeq
where $H_{S,A}$ represent symmetric ($\cT$ preserving) and
antisymmetric ($\cT$ breaking) random matrices drawn from their
respective normalized ensembles\cite{RMT}, and $C$ is a factor of
order unity which depends on the shape of the sample and the precise
nature of the chaotic scattering at the boundary\cite{univ-ham2}.
\begin{figure}
\narrowtext
\epsfxsize=2.4in\epsfysize=1.6in
\hskip 0.3in\epsfbox{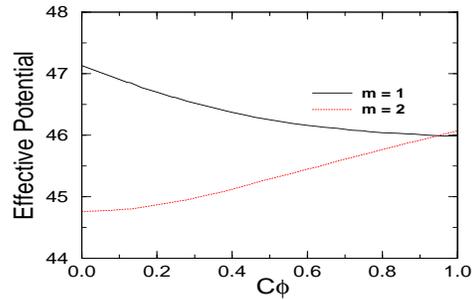}
\vskip 0.15in
\caption{The ensemble-averaged minimum of the effective
potential in units of $\Delta$ (for $g=20$ and $u_m=1.7$) as a
function of the crossover parameter $\phi$ for $m=1$ and $m=2$. }
\label{cross}
\end{figure}

Figure 2 shows the dependence of $\cE$ on the crossover parameter
$C\phi$ for $m$ even and odd, and clearly shows the diamagnetic
behaviour for $m$ even.  To see why this is special, consider what is
known about persistent currents\cite{mesoscopics-review}. In a
mesoscopic ring penetrated by a flux, the ground state energy has to
be periodic in the flux, since an integer number of flux quanta can be
gauged away.
\beq
I_{pers}(\phi)=-{\partial
F\over\partial\phi}=I_1\sin(2\pi\phi/\phi_0)+I_2\sin(4\pi\phi/\phi_0)
+\cdots
\eeq where $\phi_0=h/e$ is the flux quantum. Only the
even moments $I_{2n}$ survive disorder-averaging\cite{mesoscopics-review}.

In the noninteracting case the typical, fluctuating values of the
Fourier coefficients are (for small $n$) $I_{n,typ}\approx
{E_T/\phi_0}$ while the average is $ \langle I_{2n}\rangle\approx
{\Delta/\phi_0}$.  Experiments typically measure $\langle
I_{2n}\rangle$ and a few other low-order
harmonics\cite{persist-expt0,persist-expt0.5,persist-expt1,persist-expt1.5,persist-expt-gaas,persist-expt2}.

Interactions, when included in renormalized first-order perturbation
theory\cite{persist2}, produce $\langle I_{2}\rangle\approx
\mu^*{E_T\over\phi_0}$ where $\mu^*$ (of order $1$) is the
dimensionless Cooper-channel interaction at low energies. Thus,
interactions enhance the average persistent current, but if $\mu^*>0$
$\langle I_2\rangle$ should be paramagnetic, while if $\mu^*<0$ it
should be diamagnetic.  This prediction\cite{persist2}, while of the
same order-of-magnitude as the
experiments\cite{persist-expt0,persist-expt0.5,persist-expt1,persist-expt1.5,persist-expt-gaas,persist-expt2},
has the wrong sign. Materials that show no sign of superconductivity
(implying that $\mu^*>0$)
show\cite{persist-expt1,persist-expt1.5,persist-expt-gaas,persist-expt2}
a diamagnetic $\langle I_2\rangle$. Many explanations have been
proposed to account for this puzzle (a recent one being
Ref.\cite{far-levels}), but the question remains open, as summarized
in Ref.\cite{persist-puzzle}.  In this context a diamagnetic
persistent current of order $E_T/\phi_0$ in a model without
superconductivity is striking. The fact that our treatment is
nonperturbative in the interactions\cite{qd-us2} enables us to evade
the usual sign\cite{persist2}.  Our approach, while suggestive, is not
directly applicable to the experiments on $Au$ and $Ag$
rings\cite{persist-expt1.5,persist-expt2} since the samples are not
likely to be in the strong-coupling regime, and are not fully in the
ballistic limit (the elastic mean free path is of the same order as
the system size). On the other hand, our theory would apply directly
to ensembles of ballistic $GaAs$ rings of the type used in
Refs. \cite{persist-expt1,persist-expt-gaas}, but at stronger
coupling.

Let us now turn to $m$ odd.  The exact degeneracy of the two global
minima separated by $\pi$ in the angle $\chi$ can be seen from Figure
1, and can be proved analytically using the relation
$H_{\sigma}^*(\chi)=H_{\sigma}(\chi+\pi)$.  The two degenerate minima
are related by the time-reversal transformation $\cT$. A particular
value of $\chi$ leads to a distortion of the Fermi surface along the
direction specified by $\chi$. Under $\cT$, $\bk\to-\bk$ and a
distortion of the Fermi surface for odd $m$ maps to an inequivalent
state at $\chi+\pi$ with the same energy, since the underlying
Hamiltonian is $\cT$-invariant.  The ground state of a Hamiltonian
quantum system with a two-fold degenerate potential is the symmetric
combination of the two minima. This applies to the isolated mesoscopic
structure, whose dynamics is Hamiltonian at energy scales smaller than
$\Delta$ (for energy scales in the range $\Delta \ll
\omega\ll E_T$ the dynamics is dissipative with ohmic
dissipation, see Eq. (\ref{eff-action-s0})). The splitting between the symmetric
and antisymmetric combinations is the tunneling
amplitude between the two minima, here $\Delta e^{-g}$. The two
minima correspond to states carrying opposite persistent currents, and
are macroscopically distinguishable.

The coupling of the mesoscopic structure to the leads produces ohmic
dissipation at arbitrarily low energies. This is precisely the case of
the Caldeira-Leggett model\cite{CL} considered and solved by
Chakravarty\cite{chakravarty}, and Bray and
Moore\cite{chakravarty}. The effective action of our model at low
energies ($\w\ll\Delta$) is
\beqr
&g\int dt (V(\chi(t))+\half (d\chi/dt)^2)+\nonumber\\
&{2g\Gamma^2\langle\sigma\rangle^2\over\pi\Delta^2}\int {dt dt' \over
(t-t')^2} \sin^2\bigg({\chi-\chi'\over2}\bigg) 
\label{cl-eff-action}\eeqr
where $\chi$ is the angle of $\sigmab$ in the Mexican Hat, $V(\chi)$
is the doubly-degenerate realization-specific potential of Figure 1,
and $\Gamma$ is the level width induced by coupling to the leads. The
long-range interaction in imaginary time comes from a $|\w|$
``kinetic''term, which in turn arises from the Landau damping of
$\sigmab$ due to decay into particle-hole pairs at arbitrarily low
energies, possible because each formerly sharp level $\a$ is broadened
by coupling to the leads.

The model has a weak-dissipation phase in which the ground state is
still the symmetric superposition of the two minima, and a
strong-dissipation phase in which the particle is localized in one
minimum. The transition between the two phases\cite{chakravarty}
occurs for $g(\Gamma/\Delta)^2\langle\sigmab^2\rangle\approx1$. For
large enough $g$ even a weak coupling to the leads ($\Gamma\approx
\Delta/\sqrt{g}$, since $\langle\sigmab^2\rangle\approx1$) is
sufficient to meet this criterion, and leads to localization in one
minimum of the twofold degenerate effective potential even at zero
temperature, corresponding to a spontaneous breaking of $\cT$. The
$\cT$-breaking transition could be monitored by measuring the
peak-height statistics\cite{peak-height}.

If one turns on an external flux, one minimum moves up in energy as
the flux increases while the other moves down. The ground state (which
moves down) displays a paramagnetic persistent current of order
$E_T/\phi_0$. In the isolated dot, or in the case with weak
dissipation, one starts with a symmetric superposition of the two
minima as the zero-flux ground state. As $\phi$ increases the system
crosses over to fully $\cT$-broken dynamics when the energy difference
of the two minima is greater than their splitting in zero field, which
is $\Delta e^{-g}$.  Thus, the crossover will occur for an external
flux $\phi_{X}\approx\phi_0 e^{-g}$, as compared to the noninteracting
crossover flux $\phi\approx\phi_0/\sqrt{g}$.  For strong enough dissipation,
the ground state already breaks $\cT$, and the variation of $\cE$
contains a term first-order in $\phi$, which implies a spontaneous
persistent current at zero flux.
\begin{figure}
\narrowtext
\epsfxsize=2.4in\epsfysize=1.6in
\hskip 0.3in\epsfbox{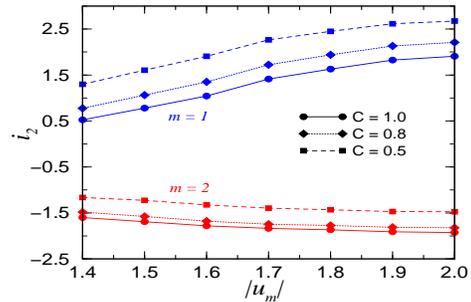}
\vskip 0.15in
\caption{The ensemble-averaged second Fourier coefficient of
the persistent current in units of $\Delta/\phi_0$ as a function of
coupling $|u_m|\ge|u_m^*|=1/\ln2$ for $m=1$ and $m=2$ and $ g = 20 $.
}
\label{i2}
\end{figure}

Figure 3 shows the ensemble-averaged second Fourier coefficient
$\langle I_2\rangle$ for $m$ even and odd. Ideally one would find a
periodic behaviour of $\cE$ with $\phi$. This cannot be captured by the
crossover Hamiltonian, but must be put in by hand. Since we do not
know the number $C$ connecting the crossover parameter to the
flux $\phi$, there is an inherent ambiguity in this procedure. We show
$\langle I_2\rangle$ for three different choices for $C$. As can be
seen, the qualitative results are unaffected.

Finite temperature produces additional interesting effects for $m$
odd. For temperatures exceeding the tunnel splitting $\Delta e^{-g}$
the superposition states are irrelevant, and one can think of the two
minima as separately thermally occupied. Because of their (near)
degeneracy, there will be a Curie-like susceptibility of $1/T$ of the
persistent current (and therefore the magnetization) to external
flux.

In summary, we have shown that some surprising signatures of the
mesoscopic Pomeranchuk regimes show up in the persistent
current. There is a diamagnetic persistent current (for $m$ even)
without any superconductivity. The $m$ odd case undergoes a
spontaneous time-reversal symmetry-breaking transition as the coupling
to the leads is increased, and displays a spontaneous persistent
current (at zero flux) in the $\cT$-broken phase.  It would be very
interesting to explore the behaviour of persistent currents in the
quantum critical regime\cite{critical-fan}, where large fluctuations
of the order parameter $\sigmab$ and finite quasiparticle
lifetime\cite{fock-loc} at low energies are expected\cite{qd-us2}.

We thank R. Shankar for illuminating discussions and E. Mucciolo for
pointing out Ref. \cite{kwon-ceperley}, and the National Science
Foundation for partial support under grants DMR 98-04983 (DH and HM)
and DMR 0071611 (GM).

\end{document}